# Unraveling Surface Chemistry of irreversible reduction of iron oxides through the Time-Resolved APXPS and Chemometrics

*Mohammed Alaoui Mansouri [1,2] ; Manoj Ghosalya [1]; Aidin Heidari [3]; Yuzhao Wang[1]; Mourad Kharbach[1,4]; Anne Hietava [3]; Mikko J. Sillanpää [2]; Ilkka Launonen [2]; Timo Fabritius [3]; Marko Huttula [1]; Samuli Urpelainen [1]*

[1] Nano and Molecular Systems Research Unit, University of Oulu, FIN-90014, Oulu, Finland

[2] Research Unit of Mathematical Sciences, University of Oulu, FIN-90014, Oulu, Finland

[3] Process Metallurgy Research Unit, Faculty of Technology, University of Oulu, Oulu, Finland

[4] Circular Economy/Sustainable Solutions, LAB University of Applied Sciences, Mukkulankatu 19, 15101, Lahti, Finland

## Abstract

This work presents a study on the application of Time-Resolved Ambient Pressure X-ray Photoelectron Spectroscopy (TR-APXPS) in association with chemometric techniques, specifically Principal Component Analysis (PCA) and Multivariate Curve Resolution with Alternating Least Squares (MCR-ALS), to investigate the surface chemistry and dynamics of Fe2O3 reduction processes. The use of TR-APXPS allows for real-time monitoring of chemical changes at the surface of Iron oxides  during reduction, providing valuable insights into the reaction mechanisms and kinetics involved.

One key challenge in analyzing TR-APXPS data is the presence of overlapping peaks and complex spectral features, which can make accurate quantification and interpretation difficult. Traditional spectral fitting methods may struggle with these complexities and result in ambiguous or inaccurate results. However, the chemometric approaches are promising tools to overcome these challenges by extracting pure spectral profiles of individual chemical species and their temporal profiles from the complex and overlapping data.

The results obtained from the TR-APXPS coupled with PCA and MCR-ALS analysis provide a detailed and precise understanding of the surface chemical changes during the Fe2O3 reduction. This includes identifying and following the formation of various intermediate species and their evolution over time, which permits later to establish correlations between surface chemistry and process conditions.

The integration of both chemometric tools in TR-APXPS data analysis not only addresses the challenges associated with complex spectral features, but also contributes to a deeper understanding of the underlying chemical changes and their dynamics. The obtained results have significant implications for process optimization, material synthesis, and tailoring of material properties for specific applications.

## 1. Introduction

X-ray Photoelectron Spectroscopy (XPS) is a fundamental technique in the analysis of surfaces, providing essential insights into the chemical composition and nature of materials by measuring the kinetic energy of electrons emitted from the surface upon X-ray bombardment [1,2]. However, conventional XPS is limited to solid samples and predominantly focuses on surfaces of semiconductors, thin films, coatings, biosensors, and dry powder pharmaceutical materials due to the requirement of a high-ultra vacuum (UHV) environment [3]. The emergence of Ambient Pressure X-ray Photoelectron Spectroscopy (APXPS) has marked a significant advancement in surface analysis. APXPS enables investigations under conditions closely resembling real-world physiochemical processes, allowing controlled analysis of materials in gases, liquids, and electrochemical cells [4]. APXPS extends the applicability of XPS to diverse fields, including catalysis, corrosion, energy storage, electronics, forensics, and biology. It provides qualitative and quantitative estimates of elemental composition and identifies chemical bonds and oxidation states on material surfaces, from lithium and beyond [5].

To enhance our understanding, time-resolved APXPS (TR-APXPS) has emerged, enabling the investigation not only of the final outcomes but also the microscopic dynamics within physiochemical processes. TR-APXPS offers insights into transient species, intermediates with short lifespans, charge transfer dynamics, and intricate reaction pathways. This technique is used to unravel complex reactions and material behaviors under dynamic conditions, highlighting fundamental processes in various applications [6].

For example, in recent studies, TR-APXPS has been combined with pulses of varying gas composition to investigate reactions [7–9]. In a study focused on heterogeneous catalysis, time resolved APXPS was used to map the catalyst structure and local gas environment simultaneously during the CO oxidation reaction on a Pd (100) model catalyst . By employing cyclic gas pulsing and software-based event-averaging through image recognition of spectral features, the study revealed the dynamics of the catalyst's response to changes in the local gas environment. The findings highlighted the reversible nature of the reaction, where the CO-covered metallic surface became highly active for a few seconds once the $O_2$:CO ratio reached a threshold, lifting the CO poisoning effect before the formation of an oxide phase.

Another study investigated the interactions between two-dimensional materials and catalytic surfaces under reaction conditions using time resolved APXPS with varying gas compositions [8].

The research demonstrated the sequential formation and removal of undercover reaction products. The results showed the rapid mixing of hydrogen into a specific structure below graphene flakes, while CO exposure led to oxygen removal from the confined space. Additionally, the study highlighted the importance of promoter chemistry by adding trace gases to the gas feed, which influenced the undercover chemistry and led to the formation of stable chemical structures.

While TR-APXPS is a powerful tool for studying surface processes and unveiling mechanistic insight of surface reactions and catalysis[10], many experimental and analysis issues need to be addressed in data acquisition step as well as in data manipulation process. TR-APXPS generates high dimensional spectroscopic data with multiple peaks corresponded to different chemical species, which are often affected by various sources of noise: background noise, instrument artifact, electronic noise, leading to the difficulty of accurate interpretation [11]. In addition, that temporal drift or fluctuation effect could further mask important information in spectra or raise the risk of misinterpreting the data. The first challenge lay on overlapping peaks. Especially when the changes between two species are minor or more than one reaction occurs at the same time during the process, this kind of signal overlap prevents determination contributions from individual chemical species and hinders deconvoluting underlying overall spectral feature. Moreover, the evolution with time of TR-APXPS spectra is usually complex and nonlinear, as chemical species do not necessarily follow uniform reaction kinetics over time [12]. This complexity introduces additional difficulties in measuring species' dynamics that can only be faced with advanced data analysis. Another challenge arises from collinearity between TR-APXPS signals. For example, synchrotron radiation may generate highly correlated spectra for different chemical species, where changes in the intensity or shape of one specie signal may influence others. As a result, interpretation can become ambiguous, potentially attributing signals to a third species because of apparent matches between measurements. This difficulty increases when investigating irreversible chemical reactions, where measurements can only be performed once, and subsequent observations require different samples, and methods such as event averaging cannot be utilized. This is the case for example in studying the reduction process of Fe2O3 reduction using H2. In such cases, alternative methods are required to capture and analyze the small changes occurring over the time from a single process run and data set [13,14].

To tackle these limitations and the complexity of such systems, chemometric techniques such as Principal Component Analysis (PCA) and Multivariate Curve Resolution-Alternating Least Squares (MCR-ALS) are essential. PCA helps dealing with large multicollinear data blocks, by reducing the dimensionality of the data set and by identifying patterns in the data; thus, it provides a conceptual sequential tool to simplify the visualization and interpretation of a given complex dataset [15–17]. MCR -ALS, based on non-negative matrix factorization, is particularly suitable for analysis of spectral datasets with overlapping signals. It can provide good spectral and concentration profiles recovery even in low signal-to-noise ratio conditions, it has also been proven to perform well in different fields including studies on catalytic reactions corrosion science and electrochemical processes where those techniques were used simultaneously to gain reliable information about surface chemistry or/and reaction dynamics [18–21].

In this work we employed TR-APXPS to pursue the reduction process of Direct Reduced Iron ore pellet (DRI), which consists mainly of hematite [22], using hydrogen, which is traditionally relies on carbon-based reductants like coke that release considerable amounts of $CO_2$. In contrast, hydrogen can reduce $Fe_2O_3$ to metallic iron while producing only water as a by-product, thus presenting a cleaner and more sustainable alternative [23]. Understanding the real-time surface transformations during hydrogen reduction such as changes in iron oxidation states is important for optimizing this process [24–26]. findings from this study might lead to inform improvements in DRI and the efficiency of production. Thus, advancing sustainable steel production.

based on the prior work [27], which focused on average data extraction methods for similar samples, it has been found that this approach captured broad trends, it lacked the resolution needed for detailed, time-resolved insights into transient species and reaction intermediates. Therefore, multivariate data analysis is employed to resolve complex spectral data and extract detailed concentration profile for each chemical species, which has been challenged to be obtained by conventional spectral fitting methods. This allows us to interpret reaction mechanisms with enhanced temporal resolution. Furthermore, our goal is extended to fit a kinetic model and estimating reaction rates to gain a more nuanced understanding of $Fe_2O_3$ reduction under hydrogen.

## 2. Ambient Pressure X-ray Photoelectron Spectroscopy measurements
### 2.1. Sample preparation and experimental procedure

The experiments were performed at the APXPS branch of the SPECIES beamline at MAX IV Laboratory [27–29]. The APXPS measurements were carried out on DRI pellets. The DRI pellets were mounted on a stainless-steel plate. Then, the sample was introduced into the high-pressure cell, which is equipped with a boron nitride heater. The boron nitride heater can sustain temperatures of up to 1400°C in UHV (Ultra-High Vacuum). However, in a hydrogen atmosphere, we achieved a maximum temperature of up to 650°C. Hydrogen gas was dosed and controlled using mass flow controllers.

The DRI sample was reduced in a hydrogen atmosphere with a flow rate of up to 5 SCCM (standard cubic centimeters per minute) with a 650°C for a duration of up to 5 hours as elucidated in figure 1 [30,31] .

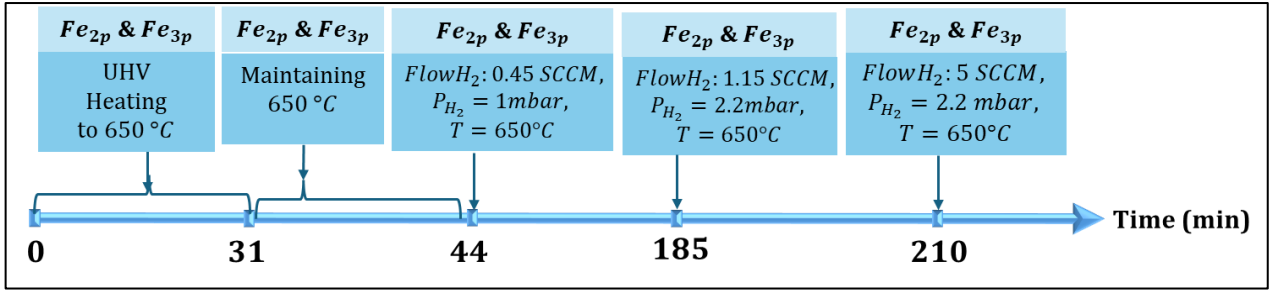

Figure 1: Timeline of Hydrogen Reduction Process for DRI Sample with Varying Flow Rates, Temperatures, and Pressures

.

## 2.2. Presentation and preprocessing the raw TR-APXPS spectra

The obtained raw spectra for DRI in Figure 2. The 3D mesh plots and their 2D projections are shown in these figures, thus allowing the visualization of how the data evolves in time during the reduction process. Importantly, raw data includes all outliers that reflect challenges during measurement. These outliers, as observed in the spectra, may be caused by several experimental factors. One common issue is sample movement during measurement, which can result in sudden changes in the spectral signal. Before conducting Principal Component Analysis (PCA) and Multivariate Curve Resolution (MCR-ALS) analysis, the raw APXPS data was normalized to a consistent intensity range for easier comparison and pattern extraction, then smoothed using a Savitzky-Golay filter to reduce random noise while preserving the shape and features of the peaks.

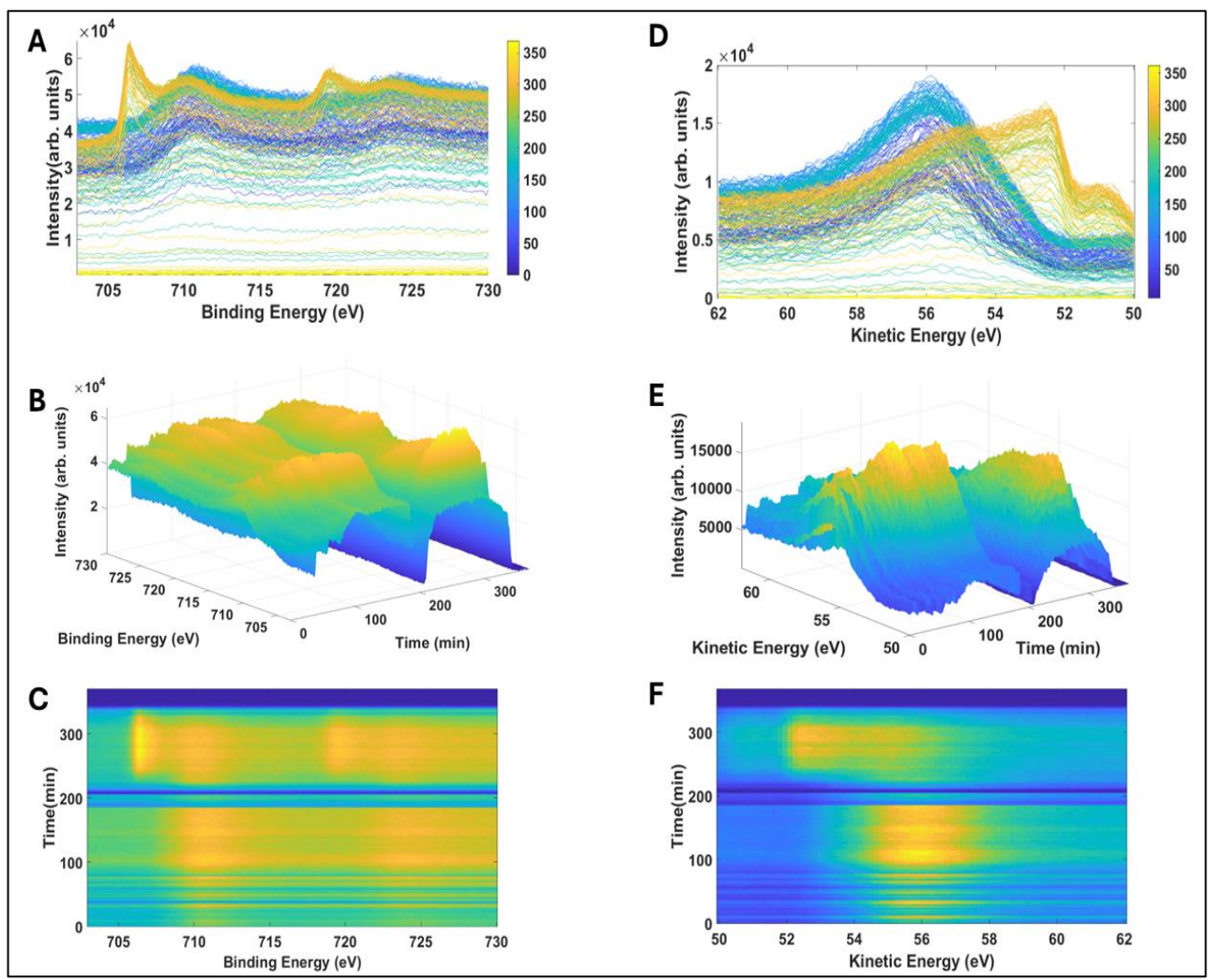

**Figure 2:** Raw spectra of DRI. (A, B and C): DRI, (D, E and F): pure powder hematite. (B) and (E) are 3D mesh plots. (C) and (F) are 2D projection

## 2.3. iron oxides: Fe 2p and 3p XPS spectra

In nature, iron oxides predominantly exist in three forms: $Fe_2O_3$ (hematite), $Fe_3O_4$ (magnetite), and FeO (wustite). $Fe_2O_3$ contains iron exclusively in the $Fe^{3+}$ oxidation state, $Fe_3O_4$ comprises a mixture of $Fe^{2+}$ and $Fe^{3+}$ oxidation states, and FeO contains iron solely in the $Fe^{2+}$ oxidation state. XPS typically employs the Fe 2p and 3p peaks to analyze the electronic structure of iron compounds. The Fe $2p_{3/2}$ peak position in $Fe_2O_3$ has been extensively studied, with reported binding energy values ranging from 709 to 711 eV [40]. Additionally, the Fe $2p_{3/2}$ peak is accompanied by satellite peaks, with the satellite for $Fe_2O_3$ occurring approximately 8 eV higher than the main peak. The satellite peak at 718.8 eV is distinctly observable and does not overlap with the main Fe $2p_{3/2}$ or Fe $2p_{1/2}$ peaks. $Fe_3O_4$, which contains both $Fe^{3+}$ and $Fe^{2+}$, exhibits Fe 2p peak positions for $Fe^{3+}$ between 709 and 711 eV and for $Fe^{2+}$ between 709 and 710 eV. A key distinction between $Fe_2O_3$ and $Fe_3O_4$ is the satellite peak at 718.8 eV, which is present in $Fe_2O_3$ but absent in $Fe_3O_4$ [33]. The FeO oxide, known as wurtzite, has binding energies of 709 eV for Fe $2p_{3/2}$ and 721.7 eV for Fe $2p_{1/2}$, with a satellite peak at 715.5 eV. However, Due to the inclusion of electrostatic interactions, spin-orbit coupling between the 2p core hole and unpaired 3d electrons of the photoionized Fe cation, and crystal field interactions, the Fe 2p peaks in all three iron oxides broaden and require fitting with multiple peaks [34].

The Fe 3p peak also exhibits spin-orbit splitting; however, the energy difference between the doublets is minimal, making them challenging to distinguish. Consequently, the spin-orbit doublet of Fe 3p peaks is conventionally treated as a single peak, with binding energies for $Fe^{3+}$, $Fe^{2+}$, and Fe at 55.6, 53.7, and 52 eV, respectively [35]. When analyzing pure samples of these iron oxides, they can be readily identified in the XPS spectra. However, complications arise when these oxides are mixed, such as during the reduction process of iron ore. The main $Fe^{3+}$ and $Fe^{2+}$ peak binding energies of $Fe_2O_3$, $Fe_3O_4$, and FeO overlap due multiple splitting of main Fe 2p peaks, making identification challenging. Nonetheless, the satellite peak for $Fe_2O_3$ at 718.8 eV and the satellite peak for FeO at 715.5 eV can assist in distinguishing the presence of $Fe_2O_3$, $Fe_3O_4$, and FeO. Despite this, manual fitting procedures remain time consuming and challenging due to the low intensity and intermixing binding energies of satellite peaks and main peak multiplets.

## 3. Multivariate data analysis

For multivariate data analysis, PCA and MCR-ALS were employed. PCA (~~Principal Component Analysis~~) is a widely used statistical method for reducing the dimensionality of a dataset while retaining the most important information. In the context of time resolved APXPS (ambient pressure Ambient Pressure X-ray Photoelectron Spectroscopy) data, PCA can be useful for understanding the variance in the data and identifying trends over time. In PCA, the data is projected onto a new set of orthogonal axes, called principal components, which are sorted in descending order of variance. The first principal component captures the most variance in the data, followed by the second, and so on. By examining the principal components, one can elucidate the underlying patterns and sources of variation in the data. To use PCA for variance explanation in time resolved APXPS data, first the data has to be preprocessed adequately to remove any effects of intensity variation. The resulting data matrix can then be subjected to PCA to obtain the principal components.

The score plot is a visualization tool that can be used to visualize the scores of the data points in the space defined by first principal components. In the context of time resolved APXPS data, the score plot can be used to identify trends in the data over time. For example, if there is a chemical reaction occurring, one might expect the data points to cluster in a particular region of the score plot corresponding to the reaction products. By examining the score plot and the loadings (which represent the contribution of each variable to each principal component), one can elucidate the sources of variation in the data and identify the most important variables for explaining that variation. This information can then be used to develop models for predicting the behavior of the system over time, or for optimizing experimental conditions to achieve desired outcomes. MCR-ALS method consists in the decomposition of the raw data matrix of evolving spectra into bilinear contributions of the pure components, mainly to concentration profile and pure spectra. Here we follow and use the MATLAB toolbox developed by Tauler et al. [36] Briefly, our data set matrix containing time resolved APXPS spectra, D, can be described as a product of pure spectra matrix (from pure species present in the reaction), S, and the respective concentration matrix, C, following the relation:

$$D = CS^T + E$$

with E the residual matrix. The method consists in wisely choosing the correct rank k of matrices C and S, then build an initial guess for one of these matrices and then refine by minimization processes with respect to the data set D. At the end, our reaction can be described dynamically by k species with pure spectra given by matrix S, evolving according to concentration profile C. In this case, PCA is used for choosing the rank and to construct the initial guess, while MCR-ALS algorithm fit C and $S^T$ according to the chemical properties and the mathematical features of each particular data set, D. Singular Value Decomposition (SVD) algorithm was used for PCA analysis with calculation of the elements to be consider in rank choice, such as, eigenvalues, score matrix and loading matrix. Constraints in MCR-ALS such as non-negativity, unimodality and closure were used to fit calculations to physical meaning of each reaction. The flexibility in where-and-how applying constraints and the capability to treat the most diverse multiset structures are the main assets of this algorithm. The art and expertise in using MCR-ALS stems from the proper selection and application of the constraints that are really fulfilled by the data set and from the ability to envision how to design and to deal with the most informative multiset structures.

For the TR-APXPS data, two data were analyzed separately using multivariate techniques, including PCA and MCR-ALS. The two data contained DRI data of the Fe 2p and 3p photoelectron spectra , respectively, in which their raw spectra were plotted in 3D mesh and in 2D projection, then they were normalized as shown in Figure 3A and Figure 3B.

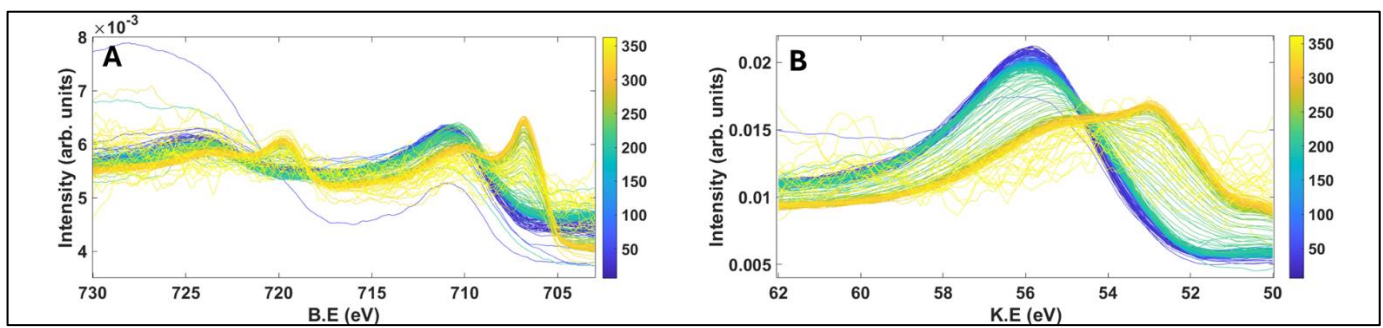

**Figure 3**: preprocessed spectra. (A): Fe2p, (B): Fe3p

Before applying PCA, the outlier spectra clearly visible in figure 3 were removed. Using Hotelling's T² vs Q residuals [37], it was possible to identify and remove spectra that have high values of Q residuals and Hotelling's T² as it is shown in Figure 4. Below we explain the PCA and MCR-ALS analyses and their results in more detail.

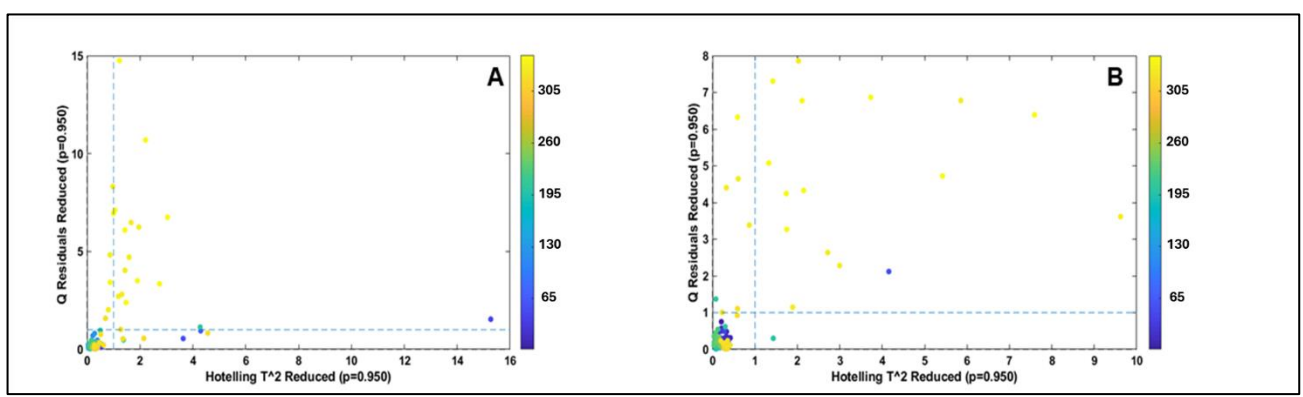

**Figure 4:** Hotelling T² vs Q residuals plot for outliers' detection in case of DRI. (A): DRI (2p). (B): DRI (3p)

### 3.1. Principal Components Analysis

Figure 4 displays the score plot of first two principal components (PC1, PC2) resulting from a PCA analysis. The x-axis represents PC1, which explains over 90% of the variance in the DRI data for both Fe 2p and 3p core levels, while y-axis represents PC2, which explains less variance than the first component. In this score plot, data points are color-coded to represent different stages of the reduction process: blue points correspond to the beginning of the process, green points indicate the middle stages, and yellow points represent the end of the process. Thus, two distinct patterns are distinguished by PC1. The color distinctions reflect the changes in binding energies as the reduction progresses. with each stage characterized by different chemical states of iron. Additionally, a slight cluster between the blue and green data points exists, with negative scores on PC2, representing samples from the beginning and middle of the process, respectively (according to Figure 5). This slight clustering is due to the differences in binding energies between hematite and the intermediate magnetite and wurtzite, as evidenced by the appearance of a satellite around 718 eV and 715 eV.

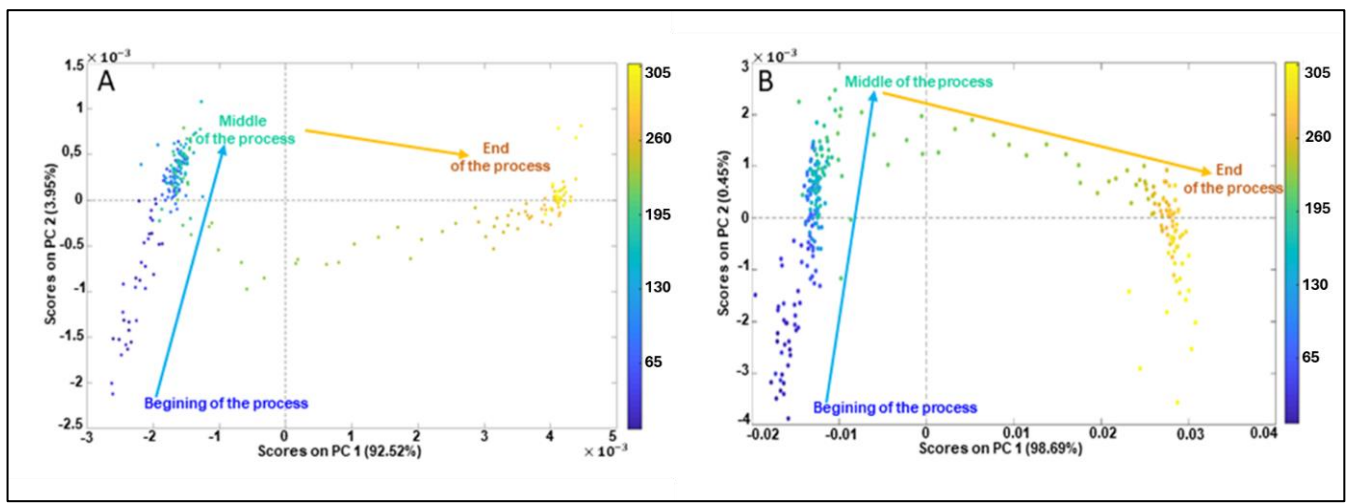

**Figure 5:** PCA Score plot of DRI: (A): Fe(2p) and (B): Fe3p,

### 3.2 Multivariate curve resolution- alternating least squares

MCR-ALS was applied, independently on Fe 2p and 3p core levels of DRI using SVD [38], that revealed the existence of four components. The SVD was followed by EFA to describe the relationship between the observed variables and the underlying factors at each time point, after which ALS optimization was performed using closure and non-negativity constraints [39]. Figure 6 shows the optimized concentration profiles for each species and their spectral identity for each core level. For Fe 2p, the first component or species corresponds to hematite due to the satellite around 718 ev, the second to magnetite, which is similar to the hematite spectral feature but characterized with the absence of the satellite, whereas the third component can be attributed to wurtzite (FeO), identified by the satellite around 715 eV, and finally the fourth component as metallic iron with distinct peak at the binding energy of 706 eV.

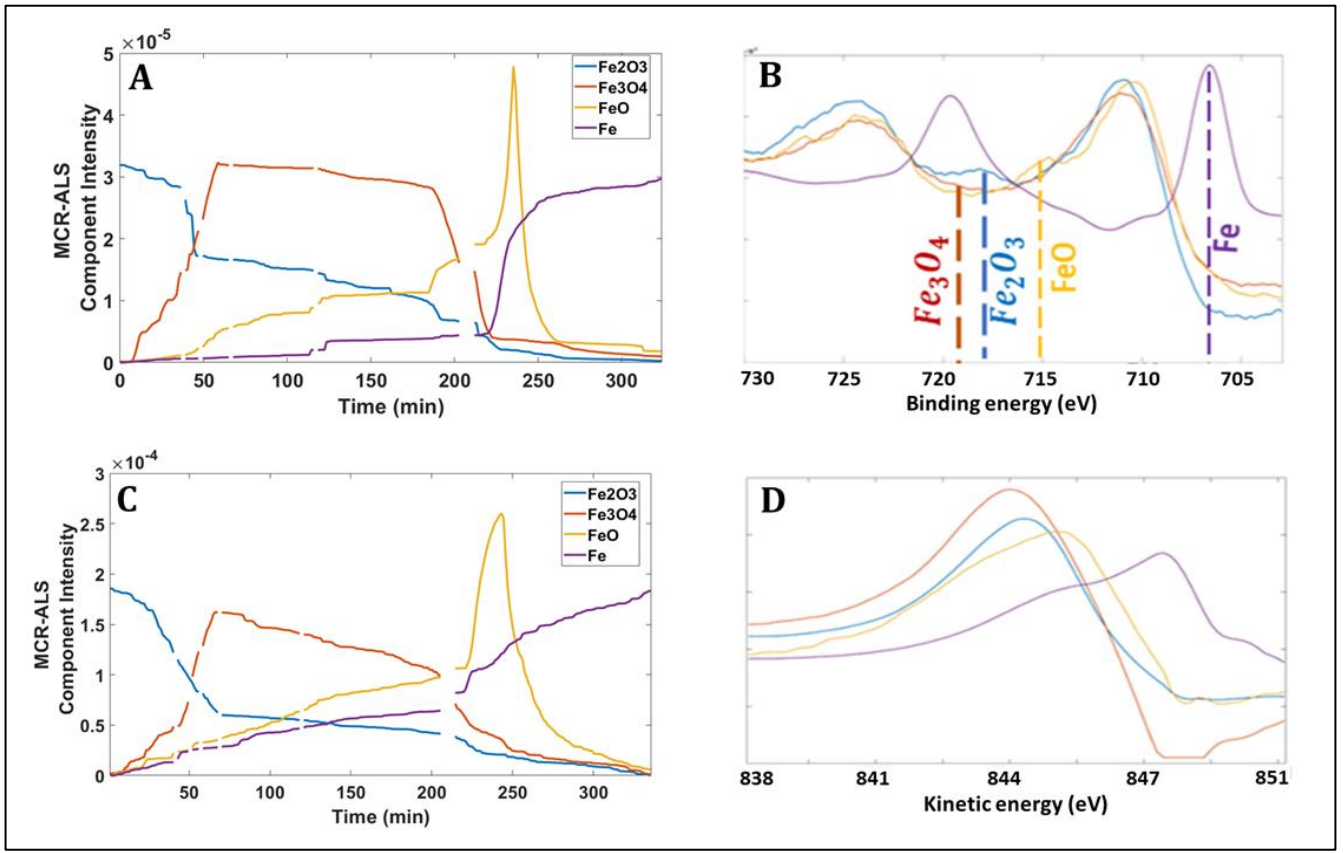

**Figure 6:** Pure spectral and concentration profiles for the four components in the resulting MCR-ALS model of DRI.
(A): concentration profile resulted by MCR of DRI (2p), (B): pure spectra profile resulted by MCR of DRI (2p),
(C): concentration profile resulted by MCR of DRI (3p), (D): pure spectra profile resulted by MCR of DRI (3p).

MCR-ALS on DRI exhibited two major challenges, whether on the Fe 2p or 3p core levels, despite the right extraction of four components representing Fe2O3, Fe3O4, FeO, and Fe.

The first challenge is caused by the co-existence of two components in the unreduced DRI, Fe2O3 and Fe3O4, in contrast to the pure powder hematite that consists only of Fe2O3 prior to reduction. These two components have only slight differences in their spectral features and can be mainly differentiated by appearance of a satellite feature that characterizes hematite and does not exist for magnetite. However; the presence of this satellite cannot be related to the absence of hematite which leads to a rank deficiency problem in applying MCR-ALS caused by similar spectral profile of the two distinct components. The problem of rank deficiency can arise when the number of determined components is lower than the number of real species that occur in the samples or measurements [40].

The second challenge is visible in the mass balance that was not respected even after applying the closure constraint. This is especially true for the concentration profile of the intermediate component corresponding to FeO, which was higher than the total concentration of all the components in the beginning of the reduction process. This was not observed only for the Fe 2p but also for the Fe 3p as shown in Figure 4. The closure constraint assumes that the mass balance is conserved and that all the species involved in the reaction have been accounted for. However, in some cases, such as for DRI in this work, It is possible that the failure of the closure constraint is caused by the different densities of the hematite (5.24 g/cm3) , magnetite (5.17 g/cm3), wustite (5.74 g/cm3) and metallic iron (7.87 g/cm3): the APXPS method probes a constant, finite area on the sample and as the sample is reduced, this area contains a changing number of Fe atoms (proportional to the density of the co-existing phases) leading to  a changing XPS signal intensity. Therefore, the probed area cannot be considered as an ideal closed system. Thus, the best alternative is to normalize by the density. The new optimized profiles of Fe2p and Fe3p is shown in figure 7.

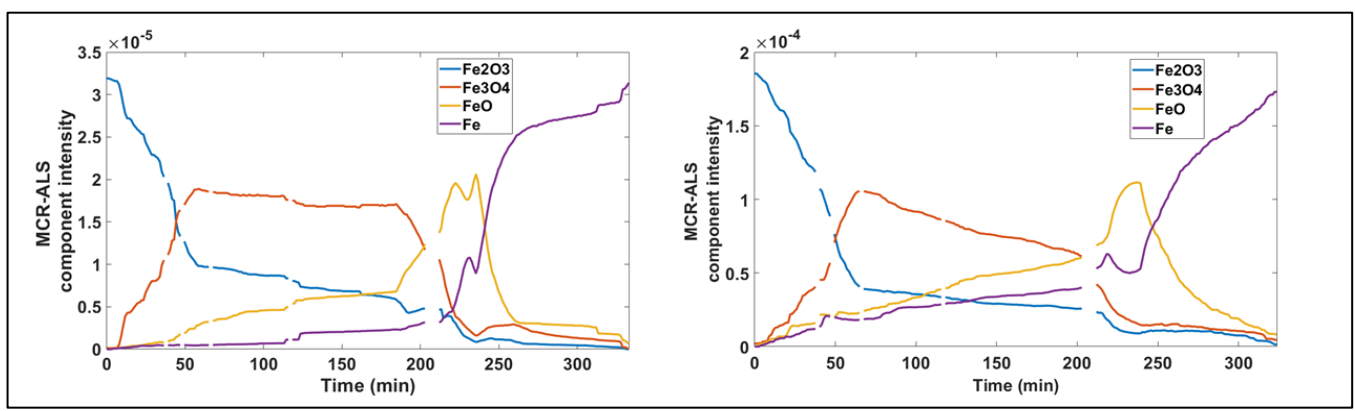

**Figure 7:** optimized concentration profiles

### 3.3.  Multivariate Analysis with MCR-ALS vs. Conventional spectral fitting

In this study, we employed MCR-ALS as a multivariate approach to analyze TR-APXPS data of the reduction process of DRI and pure powder hematite. The results obtained from MCR-ALS allowed for the simultaneous analysis of multiple overlapping spectra and kinetic profiles, providing a comprehensive understanding of the chemical species and temporal dynamics involved. The extracted concentration and pure spectra profiles were obtained without the need for explicit assumptions or constraints on the spectral contributions, making MCR-ALS a powerful and versatile method for analyzing complex datasets.

The use of MCR-ALS has clear advantages  over conventional manual data analysis in XPS studies, particularly in the analysis of degradation, depth profiling, operando, and imaging studies [41–43]. These studies have demonstrated the significance of employing advanced analytical methods to unravel the complex degradation processes observed in XPS analysis, which is relevant to our investigation of the reduction process of DRI. for example, in a recent research work, it has been discussed various chemometrics and informatics analyses applied to XPS data, including principal component analysis, cluster analysis, and most notably, multivariate curve resolution (MCR) [44]. These analyses were performed on cellulose and tartaric acid samples, leading to insights into their degradation processes. Furthermore, other multivariate spectral techniques have been employed in the analysis of X-ray photoelectron spectroscopy (XPS) data, as well as time-of-flight secondary ion mass spectrometry (ToF-SIMS) data. These techniques, such as principal component analysis (PCA) and multivariate curve resolution (MCR), offer valuable insights into complex datasets [45,46]. For example, in a study focusing on XPS and ToF-SIMS depth profiles, PCA, MCR, and pattern recognition entropy (PRE) were applied to various data sets. These analyses revealed the presence of interfaces in the films and provided indications of variations between different scan depths [47]. While our study focuses on the reduction process of DRI, the application of MCR-ALS in both studies showcases its effectiveness in extracting meaningful information from complex spectral datasets.

Compared to Conventional spectra fitting, which requires explicit assumptions or constraints on the spectral contributions, MCR-ALS offers several advantages. First, MCR-ALS can handle rank deficiency, a common issue in multivariate datasets, allowing for accurate extraction of pure spectra and concentration profiles. Second, MCR-ALS does not require prior knowledge or assumptions about the underlying spectral contributions, making it particularly useful for exploring complex datasets with overlapping spectra or kinetic profiles. Third, MCR-ALS is a computationally efficient method that can handle large datasets, making it suitable for time-resolved APXPS data analysis.

While conventional spectral fitting can also provide valuable insights, it may have limitations in complex datasets. For example, it may be prone to overfitting or require explicit constraints or assumptions on the spectral contributions, which can limit its accuracy and reliability. Conventional spectral fitting may also be computationally intensive, particularly for large datasets, and may require expertise in selecting appropriate models or fitting parameters.

However, it's important to note that MCR-ALS also has its limitations, such as the need for careful selection of constraints or regularization parameters, and the potential for uncertainties or errors in the extracted spectra or concentration profiles. Nevertheless, the insights obtained from the MCR-ALS analysis of the APXPS data in this study provide a valuable contribution to our understanding of the reduction process of DRI.

## 5. Conclusion

This study showcases the application of TR-APXPS in conjunction with chemometric techniques, specifically MCR-ALS, for investigating the surface chemistry and dynamics of Fe2O3 reduction processes. The integration of PCA and MCR-ALS in TR-APXPS data analysis addresses the challenges associated with overlapping peaks, complex spectral features, and the large dataset, enabling accurate and efficient interpretation of the chemical changes at the surface of Fe2O3 during reduction. The results yield valuable insights into the reaction mechanisms and intermediate species formation, while establishing correlations between surface chemistry and process conditions.

The success of this study highlights the potential of combining TR-APXPS with chemometric techniques for studying complex chemical systems at the molecular level, opening new possibilities for advancing our understanding of chemical processes and material properties. Future research in this area could explore the application of TR-APXPS with MCR-ALS to other complex systems and investigate the potential of other chemometric techniques for data analysis in conjunction with TR-APXPS. Additionally, the use of data fusion techniques before applying MCR-ALS could be explored as a promising avenue for further advancement in this field. Such efforts could contribute to the continued advancement of TR-APXPS and chemometric techniques for analyzing complex chemical systems and expanding their applications.

**References**

[1]  C.J. Chirayil, J. Abraham, R.K. Mishra, S.C. George, S. Thomas, Instrumental techniques for the characterization of nanoparticles, in: Thermal and Rheological Measurement Techniques for Nanomaterials Characterization, Elsevier, 2017: pp. 1–36.

[2]  E. Korin, N. Froumin, S. Cohen, Surface Analysis of Nanocomplexes by X-ray Photoelectron Spectroscopy (XPS), ACS Biomater Sci Eng 3 (2017) 882–889. https://doi.org/10.1021/acsbiomaterials.7b00040.

[3]  A.G. Shard, Chapter 4.3.1 - X-ray photoelectron spectroscopy, in: V.-D. Hodoroaba, W.E.S. Unger, A.G. Shard (Eds.), Characterization of Nanoparticles, Elsevier, 2020: pp. 349–371. https://doi.org/https://doi.org/10.1016/B978-0-12-814182-3.00019-5.

[4]  M. Borgwardt, J. Mahl, F. Roth, L. Wenthaus, F. Brauße, M. Blum, K. Schwarzburg, G. Liu, F.M. Toma, O. Gessner, Photoinduced Charge Carrier Dynamics and Electron Injection Efficiencies in Au Nanoparticle-Sensitized TiO2 Determined with Picosecond Time-Resolved X-ray Photoelectron Spectroscopy, J Phys Chem Lett 11 (2020) 5476–5481. https://doi.org/10.1021/acs.jpclett.0c00825.

[5]  J.F. Watts, J. Wolstenholme, An introduction to surface analysis by XPS and AES, John Wiley & Sons, 2019.

[6]     E. Kokkonen, M. Kaipio, H.-E. Nieminen, F. Rehman, V. Miikkulainen, M. Putkonen, M. Ritala, S. Huotari, J. Schnadt, S. Urpelainen, Ambient pressure x-ray photoelectron spectroscopy setup for synchrotron-based in situ and operando atomic layer deposition research, Review of Scientific Instruments 93 (2022) 013905. https://doi.org/10.1063/5.0076993.

[7]     J. Knudsen, T. Gallo, V. Boix, M.D. Strømsheim, G. D'Acunto, C. Goodwin, H. Wallander, S. Zhu, M. Soldemo, P. Lömker, F. Cavalca, M. Scardamaglia, D. Degerman, A. Nilsson, P. Amann, A. Shavorskiy, J. Schnadt, Stroboscopic operando spectroscopy of the dynamics in heterogeneous catalysis by event-averaging, Nat Commun 12 (2021) 6117. https://doi.org/10.1038/s41467-021-26372-y.

[8]     V. Boix, M. Scardamaglia, T. Gallo, G. D'Acunto, M.D. Strømsheim, F. Cavalca, S. Zhu, A. Shavorskiy, J. Schnadt, J. Knudsen, Following the Kinetics of Undercover Catalysis with APXPS and the Role of Hydrogen as an Intercalation Promoter, ACS Catal 12 (2022) 9897–9907. https://doi.org/10.1021/acscatal.2c00803.

[9]     M. Roger, L. Artiglia, A. Boucly, F. Buttignol, M. Agote-Arán, J.A. van Bokhoven, O. Kröcher, D. Ferri, Improving time-resolution and sensitivity of in situ X-ray photoelectron spectroscopy of a powder catalyst by modulated excitation, Chem Sci 14 (2023) 7482–7491.

[10]    A. Shavorskiy, J. Schnadt, J. Knudsen, Time Resolved Ambient Pressure X-ray Photoelectron Spectroscopy, in: Ambient Pressure Spectroscopy in Complex Chemical Environments, ACS Publications, 2021: pp. 219–248.

[11]    S. Chatterjee, B. Singh, A. Diwan, Z.R. Lee, M.H. Engelhard, J. Terry, H.D. Tolley, N.B. Gallagher, M.R. Linford, A perspective on two chemometrics tools: PCA and MCR, and introduction of a new one: Pattern recognition entropy (PRE), as applied to XPS and ToF-SIMS depth profiles of organic and inorganic materials, Appl Surf Sci 433 (2018) 994–1017. https://doi.org/https://doi.org/10.1016/j.apsusc.2017.09.210.

[12]    N. Fairley, V. Fernandez, M. Richard-Plouet, C. Guillot-Deudon, J. Walton, E. Smith, D. Flahaut, M. Greiner, M. Biesinger, S. Tougaard, D. Morgan, J. Baltrusaitis, Systematic and collaborative approach to problem solving using X-ray photoelectron spectroscopy, Applied Surface Science Advances 5 (2021) 100112. https://doi.org/https://doi.org/10.1016/j.apsadv.2021.100112.

[13]    K. Artyushkova, J.E. Fulghum, Identification of chemical components in XPS spectra and images using multivariate statistical analysis methods, J Electron Spectros Relat Phenomena 121 (2001) 33–55. https://doi.org/https://doi.org/10.1016/S0368-2048(01)00325-5.

[14]    P. Indurkar, M. Mondal, V. Kulshrestha, Highly efficient Cu-doped BTC aerogel for lead ions adsorption from aqueous solution: statistical modeling and optimization study using response surface methodology, Surfaces and Interfaces 34 (2022) 102277.

[15]    T.G. Avval, N. Gallagher, D. Morgan, P. Bargiela, N. Fairley, V. Fernandez, M.R. Linford, Practical guide on chemometrics/informatics in x-ray photoelectron spectroscopy (XPS). I. Introduction to methods useful for large or complex datasets, Journal of Vacuum Science & Technology A: Vacuum, Surfaces, and Films 40 (2022) 063206.

[16]    K.M. Mc Evoy, M.J. Genet, C.C. Dupont-Gillain, Principal component analysis: a versatile method for processing and investigation of XPS spectra, Anal Chem 80 (2008) 7226–7238.

[17]    H. Lai, J. Deng, S. Wen, Q. Liu, Elucidation of lead ions adsorption mechanism on marmatite surface by PCA-assisted ToF-SIMS, XPS and zeta potential, Miner Eng 144 (2019) 106035.

[18]   D. Oliveira  de  Souza, A. Tougerti, V. Briois, C. Lancelot, S. Cristol, Common intermediate species from reducing and activation of CoMo-based catalyst revealed via multivariate augmented system applied to time-resolved in situ XAS data, Molecular Catalysis 530 (2022) 112619. https://doi.org/https://doi.org/10.1016/j.mcat.2022.112619.

[19]   N. Veeraraghavan Srinath, H. Poelman, L. Buelens, J. Dendooven, M.-F. Reyniers, G.B. Marin, V. V Galvita, Behaviour of Platinum-Tin during CO2-assisted propane dehydrogenation: Insights from quick X-ray absorption spectroscopy, J Catal 408 (2022) 356–371. https://doi.org/https://doi.org/10.1016/j.jcat.2021.08.041.

[20]   W.H. Cassinelli, L. Martins, A.R. Passos, S.H. Pulcinelli, C. V Santilli, A. Rochet, V. Briois, Multivariate curve resolution analysis applied to time-resolved synchrotron X-ray Absorption Spectroscopy monitoring of the activation of copper alumina catalyst, Catal Today 229 (2014) 114–122. https://doi.org/https://doi.org/10.1016/j.cattod.2013.10.077.

[21]   X. Li, C. Guo, X. Jin, C. He, Q. Yao, G. Lu, Z. Dang, Mechanisms of Cr(VI) adsorption on schwertmannite under environmental disturbance: Changes in surface complex structures, J Hazard Mater 416 (2021) 125781. https://doi.org/https://doi.org/10.1016/j.jhazmat.2021.125781.

[22]   D. Guo, Y. Li, B. Cui, Z. Chen, S. Luo, B. Xiao, H. Zhu, M. Hu, Direct reduction of iron ore/biomass composite pellets using simulated biomass-derived syngas: Experimental analysis and kinetic modelling, Chemical Engineering Journal 327 (2017) 822–830. https://doi.org/https://doi.org/10.1016/j.cej.2017.06.118.

[23]   W.K. Jozwiak, E. Kaczmarek, T.P. Maniecki, W. Ignaczak, W. Maniukiewicz, Reduction behavior of iron oxides in hydrogen and carbon monoxide atmospheres, Appl Catal A Gen 326 (2007) 17–27.

[24]   D. Guo, L. Zhu, S. Guo, B. Cui, S. Luo, M. Laghari, Z. Chen, C. Ma, Y. Zhou, J. Chen, B. Xiao, M. Hu, S. Luo, Direct reduction of oxidized iron ore pellets using biomass syngas as the reducer, Fuel Processing Technology 148 (2016) 276–281. https://doi.org/https://doi.org/10.1016/j.fuproc.2016.03.009.

[25]   V. Strezov, Iron ore reduction using sawdust: Experimental analysis and kinetic modelling, Renew Energy 31 (2006) 1892–1905. https://doi.org/https://doi.org/10.1016/j.renene.2005.08.032.

[26]   M. Chu, Q. Zhao, Present status and development perspective of direct reduction and smelting reduction in China, China Metall 18 (2008) 1.

[27]   A. Heidari, M.K. Ghosalya, M.A. Mansouri, A. Heikkilä, M. Iljana, E. Kokkonen, M. Huttula, T. Fabritius, S. Urpelainen, Hydrogen reduction of iron ore pellets: A surface study using ambient pressure X-ray photoelectron spectroscopy, Int J Hydrogen Energy 83 (2024) 148–161.

[28]   S. Urpelainen, C. Såthe, W. Grizolli, M. Agåker, A.R. Head, M. Andersson, S.-W. Huang, B.N. Jensen, E. Wallén, H. Tarawneh, R. Sankari, R. Nyholm, M. Lindberg, P. Sjöblom, N. Johansson, B.N. Reinecke, M.A. Arman, L.R. Merte, J. Knudsen, J. Schnadt, J.N. Andersen, F. Hennies, The SPECIES beamline at the MAX IV Laboratory: a~facility for soft X-ray RIXS and APXPS, J Synchrotron Radiat 24 (2017) 344–353. https://doi.org/10.1107/S1600577516019056.

[29]   E. Kokkonen, F. da Silva, M.-H. Mikkelä, N. Johansson, S.-W. Huang, J.-M. Lee, M. Andersson, A. Bartalesi, B.N. Reinecke, K. Handrup, H. Tarawneh, R. Sankari, J. Knudsen, J. Schnadt, C. Såthe, S. Urpelainen, Upgrade of the SPECIES beamline at the MAX IV Laboratory, J Synchrotron Radiat 28 (2021) 588–601. https://doi.org/10.1107/S1600577521000564.

[30] Å. Rinnan, F. Van Den Berg, S.B. Engelsen, Review of the most common pre-processing techniques for near-infrared spectra, TrAC Trends in Analytical Chemistry 28 (2009) 1201–1222.

[31] V. Karki, A. Sarkar, M. Singh, G.S. Maurya, R. Kumar, A.K. Rai, S.K. Aggarwal, Comparison of spectrum normalization techniques for univariate analysis of stainless steel by laser-induced breakdown spectroscopy, Pramana 86 (2016) 1313–1327.

[32] A.P. Grosvenor, B.A. Kobe, M.C. Biesinger, N.S. McIntyre, Investigation of multiplet splitting of Fe 2p XPS spectra and bonding in iron compounds, Surface and Interface Analysis 36 (2004) 1564–1574. https://doi.org/https://doi.org/10.1002/sia.1984.

[33] M. Muhler, R. Schlögl, G. Ertl, The nature of the iron oxide-based catalyst for dehydrogenation of ethylbenzene to styrene 2. Surface chemistry of the active phase, J Catal 138 (1992) 413–444. https://doi.org/https://doi.org/10.1016/0021-9517(92)90295-S.

[34] P.S. Bagus, C.J. Nelin, C.R. Brundle, B.V. Crist, N. Lahiri, K.M. Rosso, Combined multiplet theory and experiment for the Fe 2p and 3p XPS of FeO and Fe2O3, J Chem Phys 154 (2021).

[35] T. Yamashita, P. Hayes, Analysis of XPS spectra of Fe2+ and Fe3+ ions in oxide materials, Appl Surf Sci 254 (2008) 2441–2449. https://doi.org/https://doi.org/10.1016/j.apsusc.2007.09.063.

[36] J. Jaumot, R. Gargallo, A. De Juan, R. Tauler, A graphical user-friendly interface for MCR-ALS: a new tool for multivariate curve resolution in MATLAB, Chemometrics and Intelligent Laboratory Systems 76 (2005) 101–110.

[37] K.M. Mc Evoy, M.J. Genet, C.C. Dupont-Gillain, Principal Component Analysis: A Versatile Method for Processing and Investigation of XPS Spectra, Anal Chem 80 (2008) 7226–7238. https://doi.org/10.1021/ac8005878.

[38] M.E. Wall, A. Rechtsteiner, L.M. Rocha, Singular value decomposition and principal component analysis, in: A Practical Approach to Microarray Data Analysis, Springer, 2003: pp. 91–109.

[39] A. De Juan, Y. Vander Heyden, R. Tauler, D.L. Massart, Assessment of new constraints applied to the alternating least squares method, Anal Chim Acta 346 (1997) 307–318.

[40] C. Ruckebusch, A. De Juan, L. Duponchel, J.P. Huvenne, Matrix augmentation for breaking rank-deficiency: A case study, Chemometrics and Intelligent Laboratory Systems 80 (2006) 209–214. https://doi.org/https://doi.org/10.1016/j.chemolab.2005.06.009.

[41] D.J. Morgan, S. Uthayasekaran, Revisiting degradation in the XPS analysis of polymers, Surface and Interface Analysis (2022). https://doi.org/10.1002/sia.7151.

[42] A. Erdoğan, Analysis and chemical imaging of blue inks for the investigation of document forgery by XPS, Microchemical Journal 183 (2022) 108062.

[43] F.G. de A. Dias, A.G. Veiga, A.P.A. da C.P. Gomes, M.F. da Costa, M.L.M. Rocco, Using XPS and FTIR spectroscopies to investigate polyamide 11 degradation on aging flexible risers, Polym Degrad Stab 195 (2022) 109787. https://doi.org/https://doi.org/10.1016/j.polymdegradstab.2021.109787.

[44] T.G. Avval, N. Gallagher, D. Morgan, P. Bargiela, N. Fairley, V. Fernandez, M.R. Linford, Practical guide on chemometrics/informatics in x-ray photoelectron spectroscopy (XPS). I. Introduction to methods useful for large or complex datasets, Journal of Vacuum Science & Technology A: Vacuum, Surfaces, and Films 40 (2022) 063206.

[45]    D.J. Graham, D.G. Castner, Multivariate analysis of ToF-SIMS data from multicomponent systems: the why, when, and how, Biointerphases 7 (2012) 49.

[46]    E. Saccenti, M.E. Timmerman, Approaches to sample size determination for multivariate data: Applications to PCA and PLS-DA of omics data, J Proteome Res 15 (2016) 2379–2393.

[47]    S. Chatterjee, B. Singh, A. Diwan, Z.R. Lee, M.H. Engelhard, J. Terry, H.D. Tolley, N.B. Gallagher, M.R. Linford, A perspective on two chemometrics tools: PCA and MCR, and introduction of a new one: Pattern recognition entropy (PRE), as applied to XPS and ToF-SIMS depth profiles of organic and inorganic materials, Appl Surf Sci 433 (2018) 994–1017. https://doi.org/10.1016/j.apsusc.2017.09.210.

**Figure Captions:**

**Figure 1:** Timeline of Hydrogen Reduction Process for DRI Sample with Varying Flow Rates, Temperatures, and Pressures

**Figure 2:** Raw spectra of DRI. (A, B and C): DRI, (D, E and F): pure powder hematite. (B) and (E) are 3D mesh plots. (C) and (F) are 2D projection.

**Figure 3:** Hotelling T² vs Q residuals plot for outliers' detection in case of DRI. (A): DRI (2p). (B): DRI (3p)

**Figure 4:** PCA Score plot of DRI: (A): Fe(2p) and (B): Fe3p,

**Figure 5:** Pure spectral and concentration profiles for the four components in the resulting MCR-ALS model of DRI. (A): spectra of DRI (2p), (B): concentration profile resulted by MCR of DRI (2p), (C): pure spectra profile resulted by MCR of DRI (2p), (D): spectra of DRI (3p), (E): concentration profile resulted by MCR of DRI (3p), (F): pure spectra profile resulted by MCR of DRI (3p).

**Figure 6:** Model fitting based on Fe2p spectra a) nucleation and growth models, b) diffusion models, and c) chemical reaction models.

**Figure 7:** Model fitting based on Fe3p spectra a) nucleation and growth models, b) diffusion models, and c) chemical reaction mode